\documentclass[prb,twocolumn,floats,showpacs,superscriptaddress]{revtex4}
\usepackage{graphicx,epsfig}
\usepackage{times}
\usepackage{graphics,dcolumn,bm,float}
\usepackage{amssymb,amsmath,rotate,color}

\begin{document}
\unitlength 1 cm
\newcommand{\be}{\begin{equation}}
\newcommand{\ee}{\end{equation}}
\newcommand{\bearr}{\begin{eqnarray}}
\newcommand{\eearr}{\end{eqnarray}}
\newcommand{\nn}{\nonumber}
\newcommand{\vk}{\vec k}
\newcommand{\vp}{\vec p}
\newcommand{\vq}{\vec q}
\newcommand{\vkp}{\vec {k'}}
\newcommand{\vpp}{\vec {p'}}
\newcommand{\vqp}{\vec {q'}}
\newcommand{\bk}{{\bf k}}
\newcommand{\bp}{{\bf p}}
\newcommand{\bq}{{\bf q}}
\newcommand{\br}{{\bf r}}
\newcommand{\up}{\uparrow}
\newcommand{\down}{\downarrow}
\newcommand{\fns}{\footnotesize}
\newcommand{\ns}{\normalsize}
\newcommand{\cdag}{c^{\dagger}}

\definecolor{red}{rgb}{1.0,0.0,0.0}
\definecolor{green}{rgb}{0.0,1.0,0.0}
\definecolor{blue}{rgb}{0.0,0.0,1.0}

\title{Neutral triplet Collective Mode as a new decay channel in Graphite}

\author{M. Ebrahimkhas}
\affiliation{Department of Science, Tarbiat Modares University, Tehran, Iran}
\affiliation{Department of Science, Azad University, Mahabad, Iran}

\author{S. A. Jafari{\footnote {Electronic address:
sa.jafari@cc.iut.ac.ir}}}
\affiliation{Department of Physics, Isfahan University of
Technology, Isfahan 84156-83111, Iran} 

\begin{abstract}
In an earlier work we predicted the existence of a neutral triplet collective mode in 
undoped graphene and graphite [Phys. Rev. Lett. {\bf 89} (2002) 016402]. 
In this work we study a phenomenological Hamiltonian describing the interaction of 
tight-binding electrons on honeycomb lattice with such a dispersive neutral triplet boson.
Our Hamiltonian is a generalization of the Holstein polaron problem to the case 
of triplet bosons with non-trivial dispersion all over the Brillouin zone.
This collective mode constitutes an important excitation branch which can contribute 
to the decay rate of  the electronic excitations.
The presence of such collective mode, modifies the spectral properties of electrons
in graphite and undoped graphene. 
In particular such collective mode, as will be shown in this paper, can account for 
some part of the missing decay rate in a time-domain measurement done on graphite.
\end{abstract}

\maketitle

\section{ introduction}

  Recently Novoselov and coworkers have been able to fabricate graphene, a single atomic 
layer of graphite~\cite{Novoselov}.  This discovery has brought graphene to the center of attention
of many researchers~\cite{NetoRMP}. 
The fundamental difference of the electronic spectrum of
graphene with respect to the usual metals is the existence of Fermi points around which an
effective Dirac theory describes the electronic states~\cite{semenoff}. 
The suspended graphene now can be fabricated in which the effects of impurity
and substrate is substantially reduced and one can approach the ballistic limit of 
transport with Dirac electrons~\cite{AnderiNanotech2007}.

Starting from a single layer of graphene, and adding further layers, one 
obtains, graphene multi-layers. For few layers the even-odd effects due to quantum
confinement arise~\cite{NetoRMP}. However, as the number of layers exceeds $\sim 10$, one approaches
the bulk limit, or graphite. The Dirac part of the energy dispersion of graphite 
is qualitatively similar to graphene~\cite{SaitoBook}. The only important difference between
the electronic states of graphite and graphene is the presence of small 
pockets up to $\sim 40$ meV, beyond which the Dirac description applies to 
low-energy physics of graphite as well~\cite{Kopeldhv,LanzaraGraphite}.
Ignoring such pockets which originate from the weak interlayer coupling, the electronic 
structure of bulk graphite can be approximately described by a tight binding
model on a 2D honeycomb lattice. In our approach both 
highly oriented pyrolitic graphite (HOPG) as well as undoped graphene
are treated within this model. The calculations of this paper is 
aimed to explain the life-time anomaly in HOPG, but applies to undoped
graphene as well.

The presence of Dirac points makes the nature of particle-hole excitations in graphene, 
drastically different from systems possessing extended Fermi
surface. Due to such cone like spectrum, there will be a region below the particle-hole 
continuum, where no particle-hole pairs can exist. Such a "window" does not exist in usual 
metals~\cite{baskaranjafari}. A simple random phase approximation (RPA) analysis
shows that presence of such window below the particle-hole continuum, 
provides a unique opportunity for existence of a triplet bound state of electron-hole
excitation~\cite{baskaranjafari}. An intuitive way to understand such a triplet
electron-hole bound state is to view the semi-metallic graphene from semiconducting side.
From this point of view, such collective excitation can be regarded as analogue of 
triplet excitons~\cite{jafaribaskaran}.

 In this work we focus on the life time experiment done
on HOPG sample which corresponds to
{\em undoped} graphite. The time resolved photoemission
spectroscopy (TRPES) done by Moos and coworkers~\cite{moos} on HOPG, was employed 
to measure the decay rate of quasi particles in graphite.
There are two salient features of the TRPES experiment reported by 
Moos {\em et. al.}~\cite{moos} which for convenience has been included in Fig.~\ref{decay-final.fig}:
(i) The plateau in the energy range $1-2$ eV is already a marked 
deviation from Fermi liquid prediction which was qualitatively explained in Refs.~\cite{moos,spataru},
in terms of the peculiar from of the graphite dispersion near the saddle point. 
Such a plateau has been reported in the carrier life time of doped graphene 
in angular resolved photoemission spectroscopy (ARPES) experiments as well~\cite{bostwick}
which can be understood in terms of a similar G$_0$W type of treatment~\cite{DasSarmaLifetime}.
(ii) The second important observation of the above TRPES experiment was that 
the decay rate of excitations in the whole range of energies 
over which the measurement was performed, was larger than the {\em ab-initio} 
calculation of Ref.~\cite{spataru}. This clearly means that there should be another
decay channel for quasi particles, especially in the energy range $1-2$ eV.  In the whole measurement 
range the experimentally observed decay rate is almost a factor of $2$ larger than the GW 
calculation.

Obviously the phonons cease to exist beyond $0.2$ eV, and hence can not be responsible for 
the missing decay channel in the energies reported in Ref~\cite{moos}. 
Moreover, both in HOPG and {\em undoped} graphene, there are no 
plasmons whatsoever~\cite{HwanDasSarma}. Therefore we believe this lifetime experiment 
already point to the existence of an unnoticed bosonic branch of neutral 
excitations~\cite{baskaranjafari,jafaribaskaran}. There are also other evidences
based on the Fermi velocity renormalization measurements: If one appeals to 
electron-phonon coupling to explain the experimentally observed reduction in Fermi 
velocity $v_F$ with respect to band structure prediction, one has to use an electron-phonon
coupling which is almost $\sim 5$ times larger than the density functional theory 
estimates~\cite{andrei}. 
Therefore it seems that the phonons are not enough to account for about 
$20\%$ Fermi velocity renormalization~\cite{andrei}. 
The second experimental hint for the existence of such a bosonic mode, 
is the remarkable observation of the Bose metal-insulator transition tuned by 
magnetic field~\cite{bosemetal}.

Based on the above evidences, in this paper we employ a triplet bosonic mode predicted 
in Ref.~\cite{baskaranjafari} with a 
gapless dispersion up to $\omega_{\rm max}\sim 2.1$ eV. Our model is a natural generalization
of the polaron problem, with spin-flip processes included. We generalize the momentum average (MA)
approximation developed in the context of the polaron problem by Berciu~\cite{berciu1} to 
take into account the spin-flip vertices as well as the nontrivial dispersion in the spectrum of bosons. 
First we introduce our model and the MA method.
Next we apply the MA approximation to discuss the coupling of a triplet boson to electronic
states of graphene quasi particles. The details of generalization of MA approximation to 
spin-flip processes is discussed in the appendix.

 \section{ Model and method}
We start with the Hamiltonian~(\ref{model.eqn}) describing the tight-binding electrons on honeycomb
lattice (first term), along with dispersive triplet bosons (second term)
and the interaction between electrons and bosons (third term):
\bearr
  H=\sum_{\vk,\alpha=\uparrow,\downarrow}\epsilon_{\vk}c^{\dag}_{\vk,\alpha}c_{\vk,\alpha}
  +\sum_{\vec{q},m=\pm1}\omega_{\vec{q}}S^{\dag}_{\vec{q},m}S_{\vec{q},m}\nn\\
  +g\sum_{\vk,\vec{q},m,m^\prime,\alpha,\alpha^\prime}(S^{\dag}_{\vec{q},m}
  +S_{-\vec{q},m^\prime})c^{\dag}_{\vk-\vec{q},\alpha}c_{\vk,\alpha^{\prime}},
  \label{model.eqn}
\eearr
where $\epsilon_{\vk}=\pm t\sqrt{1+\cos(\sqrt{3}k_{y}/2)\cos(k_{x}/2)+4\cos^{2}(k_{x}/2)}$ is spectrum 
of fermions for (conduction/valance) band and 
$\omega_{\vec{q}}$ describes the dispersion of spin-1 bosons~\cite{baskaranjafari}. 
Here  $c^{\dag}_{\vk,\alpha}(c_{\vk,\alpha})$ is creation (annihilation) operator for fermions 
with momentum $\vec k$ and spin $\alpha = \uparrow,\downarrow$ in either of the valence or
conduction bands, while 
$S^{\dag}_{\vec{q},m}, S_{\vec{q},m}$ are ladder operator for spin-1 bosons with momentum $\vec{q}$, 
and magnetic quantum numbers $m=\pm1,0$. In this Hamiltonian, 
$g$ is the coupling strength and describes how strongly the exchange of triplet
excitons takes place among the electrons. Estimates of a similar coupling in 
doped solid C$_{60}$ suggests $g\sim 0.3$ for those systems~\cite{BaskaranTosatti}. 
Presence of such term, favours singlet pairing under suitable 
conditions~\cite{BaskaranTosatti,Shenoy}.

   The interaction term of the Hamiltonian~(\ref{model.eqn}) describes both
spin flip ($m=\pm 1$) as well as non spin flip ($m=0$) processes.
Since non spin flip processes can exist in presence of spin-0 bosons as well,
to isolate the contribution of the spin flip processes, we focus on $m=\pm 1$
terms only. In this sector, requiring the Hamiltonian~(\ref{model.eqn}) to be
Hermitian, gives rise to the following restrictions on possible values of 
$m,m',\alpha,\alpha'$:
\be
   m=1(-1) \rightarrow m'=-1(1) \rightarrow \alpha=\downarrow(\uparrow) 
   \rightarrow \alpha'=\uparrow(\downarrow).\nn
\ee

  We use momentum average (MA) approximation to calculate the Green's function
and self-energy of system ~\cite{berciu1,berciu2} 
which yields various physical quantities such as decay rate. 
Comparison of MA and its descendants (e.g. MA(1), MA(2), etc) with other methods 
demonstrated that this method is accurate for the entire spectrum 
(both low and high energy) and for all coupling strengths and in all dimensions~\cite{berciu2}. 
This approximation was also used successfully for analysis of the effects of ripples on 
graphene sheet~\cite{berciu-graphene}.
In the following, we use MA(1) approximation, the details of which for the case 
of dispersive mode with spin-1 are derived in the appendix.

The single electron Green's function can be written as:
\bearr
   G_{\alpha,\beta}(\vk,\tau)=-i \theta( \tau)\langle 0|c_{\vk,\alpha }e^{iH\tau}c^{\dag}_{\vk,\tau}|0\rangle,
   \label{GF.eqn}
\eearr
where $\alpha,\beta$ are spin indices, and $|0\rangle$ is vacuum state. 
In the absence of bosons the free propagator is
\bearr
   G_{0}(\vk,\omega)=\frac{1}{\omega-\epsilon_{\vk}+i\eta}.
   \label{green.eqn}
\eearr
To take into account the coupling to triplet bosons, we use the equation of motion 
for $G_{\alpha,\beta}(\vk,\tau)$ to obtain (see appendix),
\be
   G_{\alpha,\beta}(\vk,\omega)=G_{0}(\vk,\omega)[\delta_{\alpha,\beta}
   +\!g\!\!\sum_{\vq_{1},m_{1}}\!\!F^{\alpha,-\beta}_{1}(\vk,\vq_{1},m_{1};\omega)],
   \label{firstgreen.eqn}
\ee
where
\be
   F^{\alpha,-\beta}_{1}(\vk,\vq_{1},m_{1};\omega)=
   \langle0|c_{\vk,\alpha}\frac{1}{\omega - \hat{H}+i\eta}
   c^{\dag}_{\vk-\vq,-\beta}S^{\dag}_{\vq,m_{1}}|0\rangle.
\ee
Here, $F_{1}$ is the amplitude for the process in which the initial state contains a
fermion and a boson, and the final states contains only a fermion with {\em opposite} spin. Hence, physically
it corresponds to the amplitude of annihilating one triplet ($\Delta m=\pm 1$) boson.
Applying again the equation of motion to $F_1$ generates hierarchy of equations
containing amplitudes with multi-boson states:
\bearr
   &&F^{\alpha,-\beta}_{1}(\vk,\vq_{1},m_{1},\omega)=\\
   &&G_{0}(\vk-\vq_{1},\omega-\omega_{\vq_{1}})[g^{2}
   \!+\!\!\!\sum_{\vq_{2},m_{2}}\!\! F^{\alpha,\beta}_{2}(\vk,\vq_{1},\vq_{2},m_{1},m_{2};\omega)].\nn
\eearr

  Although each internal vertex may contain spin flip scatterings, 
since the Hamiltonian~(\ref{model.eqn}) preserves the spin, the incoming and outgoing 
fermions must have the same spin. 
Hence the Green's function (\ref{GF.eqn}) must be diagonal with respect to the spin
indices. The rigorous proof of this is given in the appendix.
Also, by Dyson equation, the self-energy is also diagonal with respect to 
spin indices:
\bearr
   G_{\alpha,\alpha}(\vk,\omega)=[\omega-\epsilon_{\vk}-\Sigma^{\alpha,\alpha}(\vk,\omega)+i\eta]^{-1}.
   \label{finalgreen.eqn}
\eearr

   The self-energy $\Sigma^{\alpha,\alpha}(\vk,\omega)$ in MA(1) approximation is given by 
(see appendix),
\begin{widetext}
\bearr
   \Sigma^{\alpha,\alpha}(\omega)=
   \frac{g^{2}\sum_{\vk,q_{1}}G_{0}(\vk-\vq_{1},\omega-\omega_{\vq_{1}}-g^{2}A_{1}(\omega))}
   {1-g^{2}\sum_{\vk,\vq_{1}}G_{0}(\vk-\vq_{1},\omega-\omega_{\vq_{1}}-g^{2}A_{1}(\omega))(A_{2}(\omega)-A_{1}(\omega))},
   \label{sfMA1.eqn}
\eearr
\end{widetext}
where $A_1, A_2$ are functions of $\omega$, defined in the appendix.
The self-energy contains all interaction effects, and can be used to calculate 
spectral weights, decay rates, etc. in a straightforward way~\cite{mahan}.

\section{Results}
Now we are in position to derive decay rate or lifetime of quasi particles (QP) 
of HOPG/graphene in presence  of spin-1 bosonic collective mode.
There are some other decay mechanisms such as electron-hole~\cite{spataru}, 
electron-phonon~\cite{GonzalezPerfetto},
and electron-plasmon scatterings~\cite{HwanDasSarma}. In doped graphene, all 
the above mechanisms might contribute to the renormalization of QP properties. 
However, in  HOPG graphite and undoped graphene there are no plasmons to couple 
to electronic degrees of freedom.

\subsection{The decay rate}
   The imaginary part of self-energy related to life-time and decay rate of QP,
\bearr
  \frac{1}{\tau}\propto -
  \mbox{Im}[\mbox{Tr}{ \Sigma^{\alpha,\alpha}(\vk,\omega)}].
\eearr

\begin{figure}[thb]
  \begin{center}
    \includegraphics[width=9cm,height=8cm,angle=00]{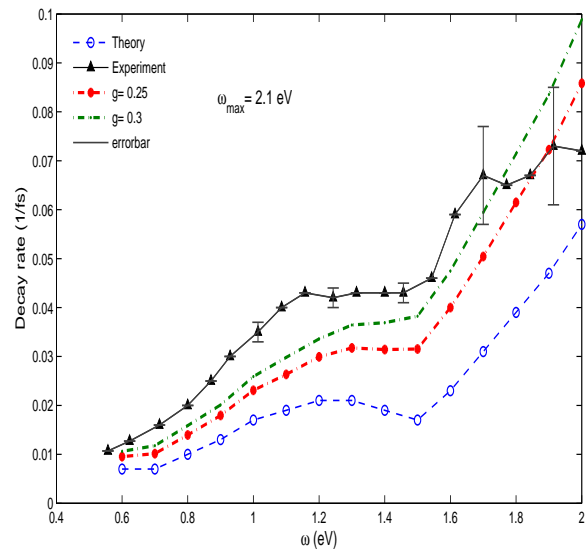}
    \vspace{-7mm}
    \caption{(Color online) Quasi particle decay rate in HOPG graphite. The triangles
    indicate TRPES measurements in Ref.~\cite{moos}, the open circles are {\em ab-initio}
    GW calculation of Ref.~\cite{spataru}. Filled circles and dashed lines
    show the total decay rate in presence of a new decay mechanism caused by
    triplet bosons for the electron-boson couplings $g=0.25,0.3$. In this calculation
    we have taken $\omega_{\rm max}=2.1$ eV. See the text for explanation.}
    \label{decay-final.fig}
  \end{center}
  \vspace{-5mm}
\end{figure}

We have numerically evaluated the integrals necessary to get the 
self-energy~(\ref{sfMA1.eqn}). In Fig.~\ref{decay-final.fig}, we
have plotted the decay rate measured in TRPES experiment of Ref.~\cite{moos}
(triangles) along with the electron-hole decay mechanisms captured within 
GW approximation (open circles)~\cite{spataru}. As can be seen in this 
figure, the decay into {\em incoherent} electron-hole pairs within GW approximation
can only account for half of the experimentally reported QP decay rate.
In this figure, we plot the total decay rate in presence of the spin flip
scatterings from a tripled bosonic mode for the coupling values 
$g=0.25$ (filled circles) and $g=0.3$ (dashed line).
The triplet bosonic collective mode has a wide dispersion between zero 
to $\omega_{\rm max}\sim 2.1$ eV. 

  As can be seen, a dispersive bosonic collective mode can account for the 
missing decay rate in HOPG graphite. The same result applies to {\em undoped} 
graphene as well. The fact that GW approximation falls a factor of two behind 
the experimentally measured decay rate, indicates that in addition to incoherent
electron-hole decay processes, there should be another decay channel provided 
by a {\em coherent} bound state of electron-hole pairs, which is what our phenomenological
Hamiltonian~(\ref{model.eqn}) describes. A simple RPA analysis
showed that such a bound state can occur in triplet channel~\cite{baskaranjafari,jafaribaskaran}.

\begin{figure}[thb]
    \vspace{0.7 cm}
    \includegraphics[width=9cm,height=6cm,angle=00]{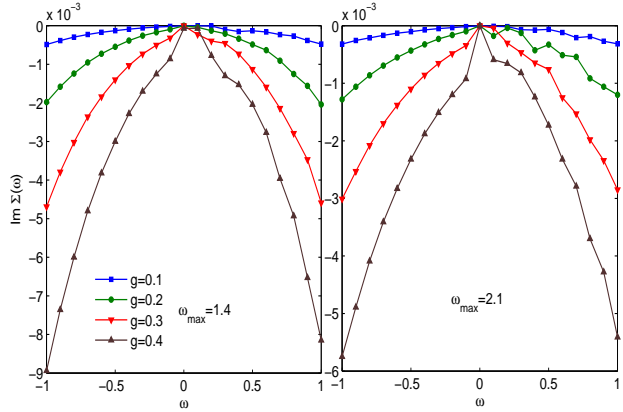}
     \caption{(Color online) Imaginary part of self-energy $vs$ energy (eV) for $\omega(q)_{max}=1.4 eV$,  $\omega(q)_{max}=2.1 eV$.} 
    \label{imsf.fig}
\end{figure}

\subsection{Dependence on $\omega_{\rm max}$}
  The dispersion of triplet bosonic mode is over a wide energy range from
zero to $\omega_{\rm max}\sim 2.1$ eV. The shape of dispersion and 
the value of $\omega_{\rm max}$ in the original work or Ref.~\cite{{baskaranjafari},{jafaribaskaran}}
is essentially controlled by the short range part of the interaction (Hubbard $U$).
It was also shown that the long-range part of the Coulomb interaction does not play
a crucial role in the dispersion of the spin-1 collective mode~\cite{jafaribaskaran}.
In the present calculation, we have fitted the dispersion relation obtained from the RPA 
analysis of Refs.~\cite{baskaranjafari,jafaribaskaran} with $\sim 10$  cosine harmonics.
over the whole Brillouin zone.

In Fig.~\ref{imsf.fig} we explore the dependence of decay rates 
on the dispersion bandwidth ($\omega_{\rm max}$). Left panel shows the
imaginary part of the self-energy for various values of the electron-boson 
coupling $g$, and for $\omega_{\rm max}=1.4$, while the left panel shows the 
same result for $\omega_{\rm max}=2.1$.  As can be seen in both panels,
by increasing the coupling strength $g$, the decay rate at a given energy scale
increases. Comparison of the left and right panels for the same values of 
$g$ shows that smaller width of dispersion ($\omega_{\rm max}$), the bosonic
mode leads to stronger spin flip scattering.
The limit $\omega_{\rm max}\to 0$ can be thought of an Einstein like phonon
mode which was studied within MA(1) in Ref~\cite{berciu-graphene}.
Smaller $\omega_{\rm max}$ in our phenomenological Hamiltonian~(\ref{model.eqn}) 
corresponds to larger $U$ in the Hamiltonian of the original 
electrons in Ref.~\cite{jafaribaskaran}. 
Hence the observation of  Fig.~\ref{imsf.fig} can be justified
as follows: In terms of the Hubbard type Hamiltonian of Ref.~\cite{baskaranjafari},
larger $U$ naturally leads to stronger decay rates. 

\subsection{Spectral function}
\begin{figure}[t]
 \begin{center}
    \vspace{0.8 cm}
    \includegraphics[width=9cm,height=8cm,angle=00]{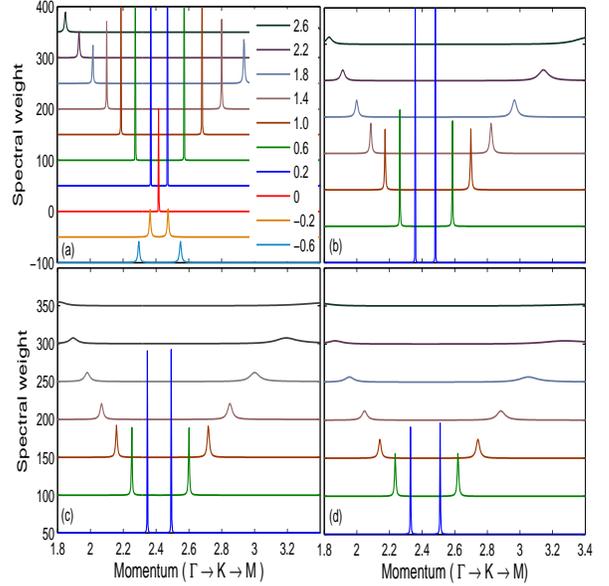}
    \caption{The spectral weight along high-symmetry cut in the $BZ$ for $\omega_{max}= 2.1 eV$ 
    and (a) $g= 0.3$ for different energy from -0.6 to 2.6 eV, (b) {\bf $g=0.6$}, (c) $g=0.8$, (d) $g=1.0$ 
    for different energy from 0.2 to 2.6 eV  }
    \label{spectral1.fig}
  \end{center}
\end{figure}

  Once we calculate the self-energy $\Sigma^{\alpha,\alpha}(\omega)$ at any approximation,
we are able to immediately  calculate the spectral weight 
$A(\vk,\omega)=-\frac{1}{2\pi}\mbox{Im}[\mbox{Tr} G_{\alpha,\alpha}(\vk,\omega)]$.
We have plotted the spectral weight along high-symmetry cut $\Gamma-K-M$ of the Brillouin zone
in Fig.~\ref{spectral1.fig}. We have plotted the spectral weight for different energies.
Panels (a)-(d) correspond to different values of $g$ as indicated in the figure caption.

  The first point to notice in all panels is that the cone like dispersion 
of the Dirac electrons remains quite robust against increase in the electron-boson
coupling $g$. To see this more transparently, in panel (a) we have plotted some negative
energy spectral functions as well. As can be seen in panel (d), large values of
coupling $g\sim 1$ lead to a remarkable broadening in the quasi particle peaks. 
Direct comparison with ARPES experiments on graphene indicates that $g$ can not
be as large as $g\sim 1$. 

   Negative energy plots of panel (a) indicates that there is an asymmetry
between the positive energy states and the negative energy states. This is
natural, as the collective mode is an excitation and does not carry negative
energies.

\section{conclusion}
In this work we considered a phenomenological Hamiltonian containing a 
neutral spin-1 collective mode as a new bosonic branch of excitaitons
predicted to exist in HOPG and undoped graphene~\cite{baskaranjafari}.
Employing the momentum average self-energy we showed that such a coherent
particle-hope bound state in triplet channel can account for substantial part
of the missing decay rate in TRPES experiment of Ref.~\cite{moos} in HOPG.
Another supporting evidence for existence of such a spin-1 collective mode
which is a natural generalization of triplet excitations of ordinary 
semiconductors to the case of semi metallic HOPG comes from the 
downward renormalization of $v_F$~\cite{andrei}. Apparently phonons fail
to account for such renormalization. Moreover, the remarkable observation
of Bose metal-insulator transition tuned by magnetic field in HOPG~\cite{bosemetal},
might indicate that there such spin excitation branch can have interesting
consequences for the behavior of HOPG and graphene in magnetic fields.

\section{acknowledgement}
We wish to thank M.R. Abolhassani, Y. Kopelevich, M. Berciu and G. Baskaran for 
comments and suggestions. S.A.J. was supported by the Vice Chancellor
for Research Affairs of the Isfahan University of Technology, and the
National Elite Foundation (NEF) of Iran.

\appendix
\section{Generalization of MA(1) for spin-flip Hamiltonians}
We start by writing Eqn.~(\ref{firstgreen.eqn}) with explicit spin indices. 
The matrix elements of Green's function become,
\bearr
   G_{\up,\up}(\vk,\omega)=&\!\!G_{0}(\vk,\omega)[1\!+\!g\!\sum_{\vk,\vq_{1}}\!F^{\up,\down}_{1}(\vk,\vq_{1},+1;\omega)]
   \label{1.a}\\
   G_{\down,\down}(\vk,\omega)=&\!\!G_{0}(\vk,\omega)[1\!+\!g\!\sum_{\vk,\vq_{1}}\!F^{\down,\up}_{1}(\vk,\vq_{1},-1;\omega)]
   \label{2.a}\\
   G_{\up,\down}(k,\omega)=&\!\!\!G_{0}(k,\omega)[g\sum_{k,q_{1}}F^{\up,\up}_{1}(k,q_{1},-1;\omega)]
   \label{3.a}\\
   G_{\down,\up}(\vk,\omega)=&\!\!\!G_{0}(\vk,\omega)[g\sum_{\vk,\vq_{1}}F^{\down,\down}_{1}(\vk,\vq_{1},+1;\omega)].
   \label{4.a}
\eearr
\begin{figure}[t]
 \begin{center}
    \vspace{0.8 cm}
    \includegraphics[width=9cm,height=3cm,angle=00]{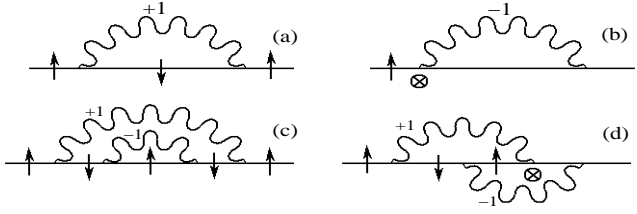}
    \caption{ First and second order Feynman diagrams for the scattering vertex 
    from a spin-1 collective mode. At each vertex the spin of electron is flipped.
    Therefore, the incoming and outgoing spins end up to be identical as in diagrams (a), (c).
    However, if we insist to have the spin of outgoing state to be opposite to that of
    incoming state, some vertices (denoted by $\otimes$) will not do flip the spin; that is they will not
    correspond to scattering from a spin-1 collective mode, as in diagrams (b), (d).} 
    \label{Feynman_diagram.fig}
  \end{center}
\end{figure}

As can be intuitively seen in Fig~\ref{Feynman_diagram.fig},
the non diagonal element of Green's function should be zero. To see this more
systematically, one writes the one bosons Green's function as,
\bearr
  \sum_{\vk,\vq_{1}}F^{\up,\up}_{1}(\vk,\vq_{1},-1;\omega)=\frac{G_{\down,\up}(\vk,\omega)}{gG_{0}(\vk,\omega)}.
  \label{5.a}
\eearr
Repeating the equation of motion we obtain the two boson amplitude:
\bearr
   &&\sum_{\vq_{1},\vq_{2}}F^{\up,\down}_{2}(\vk,\vq_{1},\vq_{2},-1,+1;\omega)= \\
   &&\frac{G_{\down,\up}(\vk,\omega)}{gG_{0}(\vk,\omega)}\left[\frac{1}{gG_{0}(\vk,\omega)}-gG_{0}(\vk-\vq,\omega-\omega(\vq))\right]\nn
   \label{6.a}
\eearr
Finally for order $N+1$, we obtain for even $N$:
\bearr
   &\sum_{\vk,\vq_{1},\vq_{2},...,\vq_{N+1}}F^{\alpha,\alpha}_{N+1}(\vk,\vq_{1},\vq_{2},...,+1,-1,...;\omega)=\nn\\
   &A(\vk,\vq_{1},\vq_{2},...,+1,-1,...;\omega)G_{\up,\down}(\vk,\omega)=0,
\eearr
and for odd $N$:
\bearr
   &\sum_{\vk,\vq_{1},\vq_{2},...,\vq_{N+1}}F^{\up,\down}_{N+1}(\vk,\vq_{1},\vq_{2},...,+1,-1,...;\omega)=\nn\\
   &B(\vk,\vq_{1},\vq_{2},...,+1,-1,...;\omega)G_{\up,\down}(\vk,\omega)=0.
\eearr
The $A,B$ coefficients for various orders can be seen by inspection to be
non zero. This proves that the spin off-diagonal components of the Green's
function are zero. This can be seen intuitively in Fig.~\ref{Feynman_diagram.fig}.

To proceed further, we define a modified form of the bosonic Green's 
function in MA(1) approximation as:
\bearr
   &&f^{\alpha,-\alpha}_{n}(\vk,\vq_{1},...,\vq_{n},...+1,-1,...;\omega)=\nn\\
   &&\frac{g^{n}F^{\alpha,(-1^{n})\alpha}_{n}(\vk,\vq_{1},...,\vq_{n},...+1,-1,...;\omega)}{G_{\alpha,\alpha}(\vk,\omega)}.
   \label{7.a}
\eearr
By inserting this in Eq.(\ref{firstgreen.eqn}) one finds,
\bearr
   &&G_{\alpha,\alpha}(\vk,\omega)=\nn\\
&&\!\!\!\!\!\!\!\!\!\!G_{0}(\vk,\omega)[1+g\sum_{\vk,\vq_{1}}f^{\alpha,-\alpha}_{1}(\vk,\vq_{1},1;\omega)G_{\alpha,\alpha}(\vk,\omega)].
   \label{8.a}
\eearr
where the matrix form of the Green's function is:
\bearr
   G(k,\omega)=\left(
   \begin{array}{cc}
      G_{\up,\up}(\vk,\omega) & 0\\
      0 & G_{\down,\down}(k,\omega)
   \end{array}\right).
  \eearr
Dyson's equation,
\bearr
  G_{\alpha,\alpha}(\vk,\omega)=[\omega-\epsilon_{\vk}-\Sigma^{\alpha,\alpha}(\vk,\omega)+i\eta]^{-1},
\eearr
give the  spin diagonal self-energy as,
\bearr
   \Sigma^{\alpha,\alpha}(\omega)=\sum_{\vk,\vq_{1}}f^{\alpha,-\alpha}_{1}(\vk,\vq_{1},+1;\omega).
\eearr
So the MA(1) self-energy is diagonal and can be casted into the final form
given by Eq.~(\ref{sfMA1.eqn}), where
\bearr
   A_{1}(\omega) &=& \frac{\bar{g}_{0,2}(\omega)}{1-\frac{2\bar{g}_{0,3}(\omega)\bar{g}_{0,2}(\omega)}{1-\ldots}},\nn\\
   A_{2}(\omega) &=& \frac{2\bar{g}_{0,2}(\omega)}{1-\frac{3\bar{g}_{0,3}(\omega)\bar{g}_{0,2}(\omega)}{1-\ldots}},\\
   \bar{g}_{0,n}(\omega) &=& \sum_{\vk,\vq_{1},...,\vq_{n}}G_{0}(\vk-\sum_{i}\vq_{i},\omega-\sum_{i}\omega_{\vq_{i}}).\nn
\eearr

\end{document}